\documentclass[lettersize,journal]{IEEEtran}
\usepackage{amsmath,amsfonts}
\usepackage{array}
\usepackage{multirow}
\usepackage{textcomp}
\usepackage{stfloats}
\usepackage{url}
\usepackage{verbatim}
\usepackage{graphicx}
\usepackage{cite}
\usepackage{caption} 
\usepackage{amsmath}
\usepackage{xspace}
\usepackage{fixltx2e}
\usepackage{subcaption}
\usepackage{algorithm}
\usepackage{algpseudocode}
\usepackage{enumitem}

\graphicspath{{Figure/}}
\algnewcommand{\algorithmicor}{\textbf{ or }}

\begin{document}

\title{Optimization of Probabilistic Constellation Shaping for Optical OFDM Systems with Clipping Distortion}

\author{Thanh~V.~Pham,~\IEEEmembership{Senior Member,~IEEE}
        and Susumu Ishihara, ~\IEEEmembership{Member,~IEEE}
\thanks{Thanh V. Pham and Susumu Ishihara are with the Department of Mathematical and Systems Engineering, Shizuoka University, Shizuoka, Japan (e-mail: pham.van.thanh@shizuoka.ac.jp, ishihara.susumu@shizuoka.ac.jp).}}
\maketitle
\begin{abstract}
Optical orthogonal frequency-division multiplexing (OFDM) and probabilistic constellation shaping (PCS) have emerged as powerful techniques to enhance the performance of optical wireless communications (OWC) systems. While PCS improves spectral efficiency and adaptability, we show that its integration with optical OFDM can inadvertently increase the peak-to-average power ratio (PAPR) of the signal, exacerbating clipping distortion due to signal clipping. This letter investigates the impact of PCS on the PAPR of direct current-biased optical OFDM (DCO-OFDM) waveforms and proposes an optimization of PCS that maximizes channel capacity, considering clipping distortion. The optimization problem is shown to be complex and non-convex. We thus present a suboptimal yet efficient solving approach based on projected gradient descent to solve the problem. Simulation results demonstrate the superiority of the proposed approach over the conventional uniform signaling, particularly under severe clipping distortion conditions.
\end{abstract}
\vspace{-0.5cm}
\section{Introduction}
The demand for ultra-high data-rate requirements of 6G wireless communications, coupled with the increasing scarcity of the radio-frequency (RF) spectrum, has motivated research and development of optical wireless communications (OWC) technology. Operating at vast and unlicensed spectrum, OWC systems are capable of offering multi-Gbps transmission rates while not interfering with RF signals \cite{Celik2023}.

OWC systems often suffer from inter-symbol interference (ISI) caused by multipath transmissions (especially due to reflections off the walls in indoor scenarios). Traditionally, in RF systems, ISI can be effectively mitigated by employing orthogonal frequency division multiplexing (OFDM). However, conventional OFDM can not be directly applied to OWC systems, as it generates complex-valued and bipolar signals, while OWC systems employing intensity modulation and direct detection (IM/DD) require real-valued, non-negative signals. To bridge this gap, several OFDM variants tailored for optical systems have been proposed, including direct current biased optical OFDM (DCO-OFDM) and asymmetrically clipped optical OFDM (ACO-OFDM)\cite{SurveyACODCO}. These optical OFDM schemes apply Hermitian symmetry to ensure real-valued outputs before the inverse fast Fourier transform (IFFT), followed by techniques such as DC biasing and clipping to produce non-negative time-domain signals.

Aside from OFDM, constellation shaping, which optimizes the location and/or the occurrence probability of the constellation points with respect to the channel condition, is a promising technique to improve the performance of communications systems. Optimizations of the location and occurrence probability of the constellation points are respectively known as geometric constellation shaping (GCS) and probabilistic constellation shaping (PCS). While it has been proven that GCS is not practical for commercial systems, PCS is receiving considerable interest recently, particularly in optical fiber communications (OFC), due to the development of the probabilistic amplitude shaping (PAS) architecture \cite{Cho2019}, which significantly simplifies the integration of shaping and coding.   

Inspired by the success in OFC systems, over the past few years, several studies have investigated the use of PCS for OWC, including free-space optical (FSO) and visible light communications (VLC) systems. For FSO, Elzanaty \textit{et al.} introduced a PCS-enhanced adaptive coded modulation scheme tailored for FSO backhauling using intensity modulation/direct detection (IM/DD) \cite{Elzanaty2020}. Subsequent works then explored the performance of PCS in VLC systems, for single-user \cite{Gutema2020, Kafizov2022, Pham2025} and multi-user \cite{Kafizov2024, Nguyen2025} scenarios. Most previous works considered single-carrier modulations such as on-off keying (OOK) and pulse amplitude modulation (PAM), whereas only \cite{Gutema2020} investigated PCS in conjunction with OFDM signaling. Similar to the case of single-carrier modulations, PCS proves advantageous in improving the performance of multi-carrier systems. However, it is important to note that the impact of PCS on the peak-to-average power ratio (PAPR) of OFDM waveforms has not been taken into consideration in \cite{Gutema2020}. \textit{To the best of our knowledge, there has been no investigation regarding this issue so far. In this study, we argue that, without a proper optimization, PCS can negatively influence the PAPR statistics of OFDM signaling, at least in the case of DCO-OFDM}. Specifically, extensive simulations in the later part of the paper show that compared with uniform signaling, PCS statistically incurs higher PAPR. In OWC systems, due to the limited dynamic linear range of the optical front-end devices and the requirement of nonnegative signals induced by intensity modulation, OFDM signals should be properly clipped, which results in clipping distortion. Since the severity of clipping distortion is proportional to the PAPR, employing PCS may lead to an increased clipping distortion. 

The purpose of this work is to design an optimization of PCS for DCO-OFDM systems considering clipping distortion. Our main contributions are summarized as follows.
\begin{itemize}[leftmargin=*]
    \item  Through simulations, we show that compared with uniform signaling, PCS statically incurs a higher PAPR of DCO-OFDM waveforms, thus potentially leading to a worse clipping distortion. 
    \item We formulate a PCS optimization problem whose objective is to maximize the signal-to-noise-plus-distortion ratio of a clipped DCO-OFDM signal. The problem is shown to be complex and nonconvex, making it generally difficult to solve optimally. Thus, we present a computationally efficient suboptimal solution based on the projected gradient descent (PGD) and nested bisection methods to solve the problem.   
\end{itemize}
\section{System model}
\subsection{DCO-OFDM}
\begin{figure}[t]
    \centering
    \includegraphics[scale = 0.26]{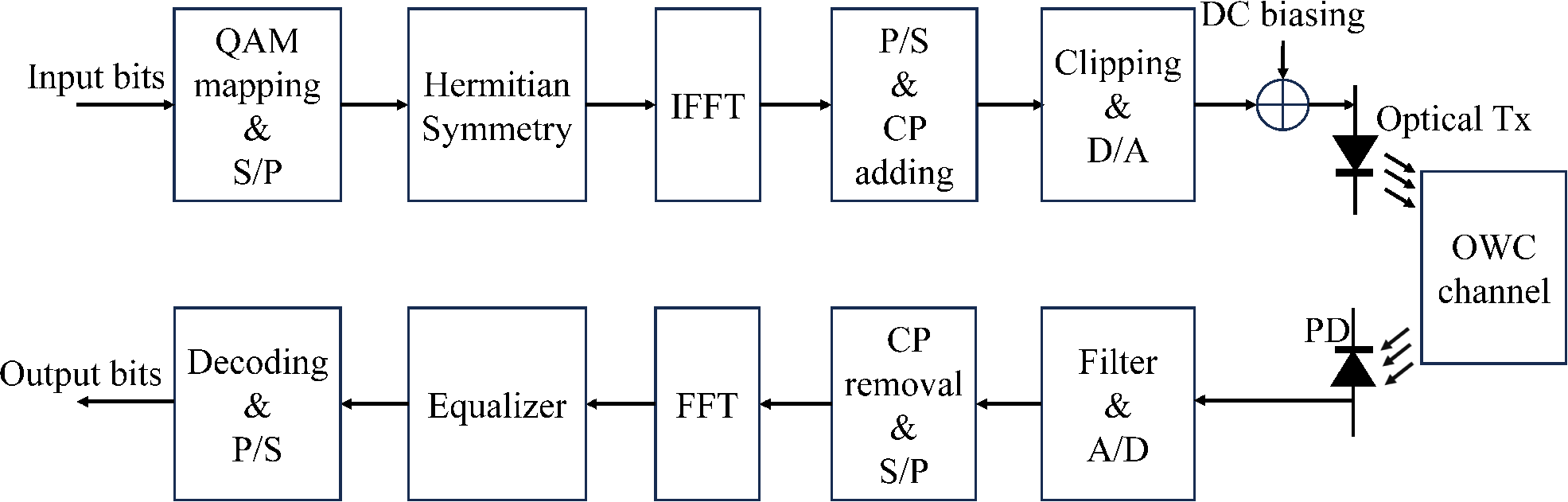}
    \caption{A block diagram of DCO-OFDM systems.}
    \label{fig:DCO-OFDM}
    \vspace{-0.4cm}
\end{figure}

In this paper, we consider DCO-OFDM systems employing $M$-ary QAM signaling with $N$ subcarriers ($N >0$ and is even)\footnote{The same analysis can be readily applied to other optical OFDM variants, such as ACO-OFDM.}. As illustrated in Fig.~\ref{fig:DCO-OFDM}, at the transmitting side, the input bits are mapped onto a probabilistic QA constellation.  OWC systems often use intensity modulation/direct detection (IM/DD), which requires that the time-domain signal be real and non-negative. To fulfill the requirement of real-valued time-domain signals, Hermitian symmetry is imposed on the OFDM symbols before performing the Inverse Fast Fourier Transform (IFFT). Let $\mathbf{X} = \big[X[0], X[1], \ldots, X[N-1]\big]$ denote the frequency-domain vector of an OFDM symbol. The Hermitian symmetry on $\mathbf{X}$ implies $X[0] = X[N/2] = 0$ and 
\begin{equation}
X[k] = X^*[N-k], \quad \text{for } 1 \leq k \leq \frac{N}{2} - 1,
\label{equation1}
\end{equation}
where $X^*[N-k]$ denotes the conjugate transpose of $X[N-k]$. 
The time-domain signal after IFFT is  given by
\begin{equation}
x[n] = \frac{1}{\sqrt{N}} \sum_{k=0}^{N-1} X[k]e^{j2\pi kn/N}, \quad n = 0, 1, \ldots, N-1.
\end{equation}
A cyclic prefix (CP) is then added to the signal for combating inter-symbol interference (ISI) and preserving orthogonality between subcarriers in the presence of multipath fading. 
In OWC systems, the optical transmitter (e.g., an LED or a laser) often exhibits a limited linear dynamic range over which the output optical power is linearly proportional to the input current. To ensure an efficient operation of the optical device, the time domain signal $x[n]$ needs to be clipped. Moreover, the clipping process should be performed in the digital domain to efficiently utilize the dynamic range of the digital-to-analog (D/A) converter \cite{Dimitrov2012}. After the conversion, the clipped analog signal is biased by a DC bias to generate a non-negative signal.  

At the receiving side, the transmitted optical signal is converted back to an electrical signal via a photodiode (PD), which is then transformed to the digital domain using an analog-to-digital (A/D) converter. After the CP removal, the Fast Fourier Transform (FFT) is performed to convert the time-domain signal back into the frequency domain, which is used for demodulation. 
\subsection{Impact of PCS on PAPR}
In multi-carrier modulation schemes such as OFDM, the summation of many subcarriers occasionally results in large peaks. This inevitably leads to a signal waveform having high PAPR, which is defined by
\begin{align}
    \text{PAPR}\left\{x[n]\right\} = \frac{\underset{0 \leq n \leq N - 1}{\text{max}} \left|x[n]\right|^2}{\mathbb{E}\left[|x[n]|^2\right]}. 
\end{align}
In the case of radio frequency (RF) systems, high PAPR signal waveforms cause high power amplifiers (HPAs) to operate in their nonlinear region, resulting in nonlinear signal distortions and reduced power efficiency. On the other hand, in the case of optical systems, the limited linear dynamic range of optical transmitters and the nonnegative signal constraint often require high PAPR signals to be clipped, which gives rise to clipping distortion.  

The use of constellation shaping to reduce the PAPR in OFDM systems has been investigated in several works \cite{Mobasher2006,Sterian2006}. Nonetheless, it is worth mentioning that previous studies examined GCS rather than PCS. While it has been thoroughly verified that PCS is beneficial to improving spectral efficiency, to the best of our knowledge, how PCS influences the PAPR statistics of OFDM waveforms has not been investigated. In Fig.~\ref{fig:constellation}, we numerically evaluate the complementary cumulative distribution function (CCDF) of PAPR for DCO-OFDM with uniform signaling and PCS. The CCDF, which is defined by $
\text{CCDF} = \Pr(\text{PAPR} \geq \text{PAPR}_0)
$ where $\text{PAPR}_0$ is a predefined PAPR threshold, is often used to asses the severity of PAPR. Our simulations examine DCO-OFDM systems assuming 4-, 8-, 16-, and 32-QAM constellations with $N = 64$ and 128 subcarriers, CP length of $N/4$. Without loss of generality, assume that the average symbol power is normalized to unity, i.e., $\mathbb{E}\left[\left|X[k]\right|^2\right] = 1, ~\forall k \neq 0, N/2$. For each constellation, 100,0000 symbols are generated to calculate the CCDF. The CCDF results in the case of PCS are obtained by averaging over 1000 randomly generated constellation distributions.    

Over the PAPR threshold ranging from 5 to 20 dB, simulation results clearly show that the CCDF of PAPR in the case of PCS is higher than that in the case of uniform signaling, particularly in the low modulation order regime (i.e., $M \leq 16$). {\textit{It is thus evident that while nonuniform constellation distributions due to PCS can improve the energy efficiency, they may also increase the PAPR of OFDM waveforms, leading to worse clipping distortions}}. It is, therefore, imperative to investigate the optimization of PCS for DCO-OFDM systems considering clipping distortion. 
\begin{figure}[ht]
    \centering
    \begin{subfigure}{0.24\textwidth}
        \centering
        \includegraphics[width=\linewidth,height = 0.7\linewidth]{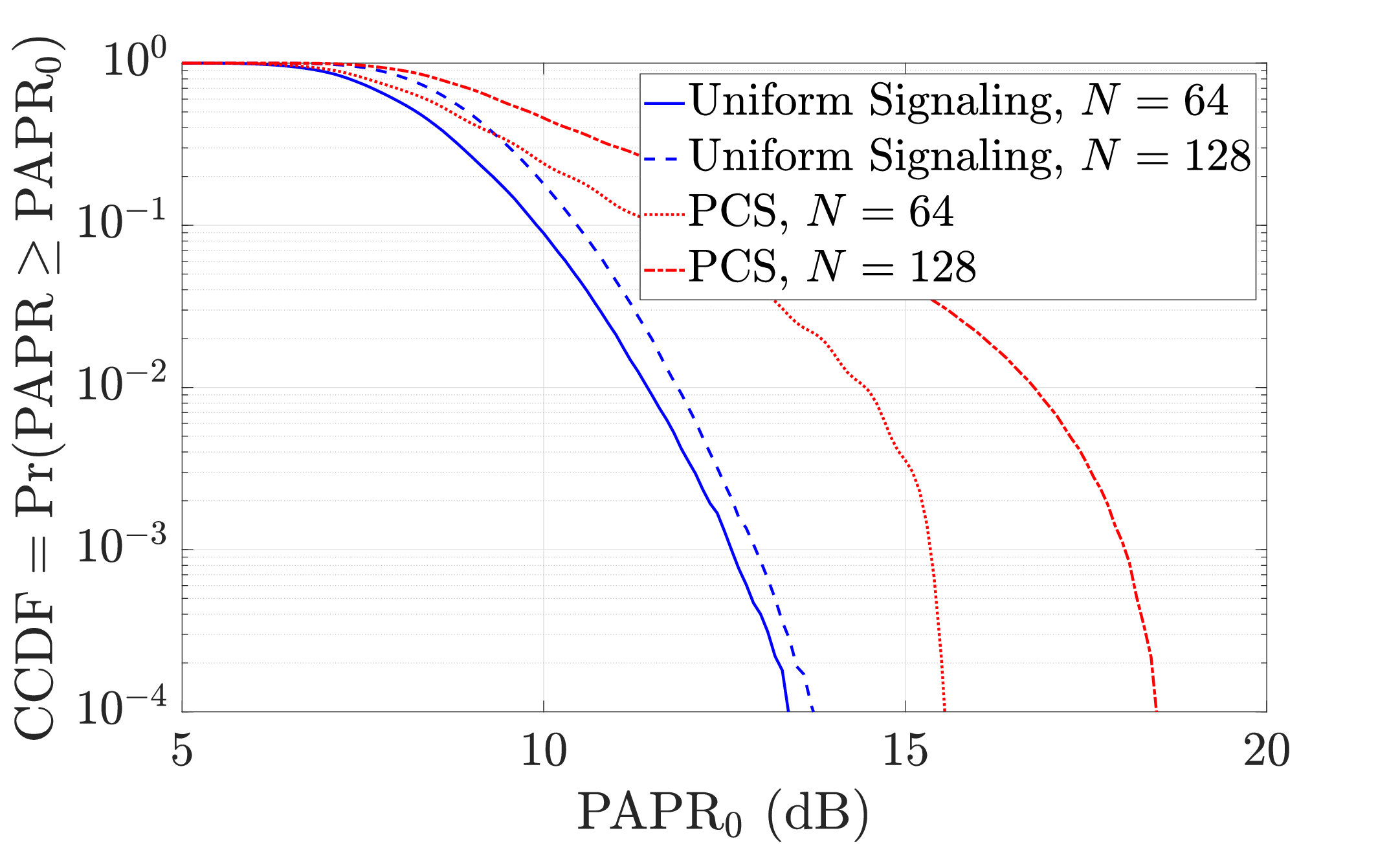}
        \caption{4-QAM.}
        \label{fig:4QAM}
    \end{subfigure}
    \begin{subfigure}{0.24\textwidth}
        \centering
        \includegraphics[width=\linewidth,height = 0.7\linewidth]{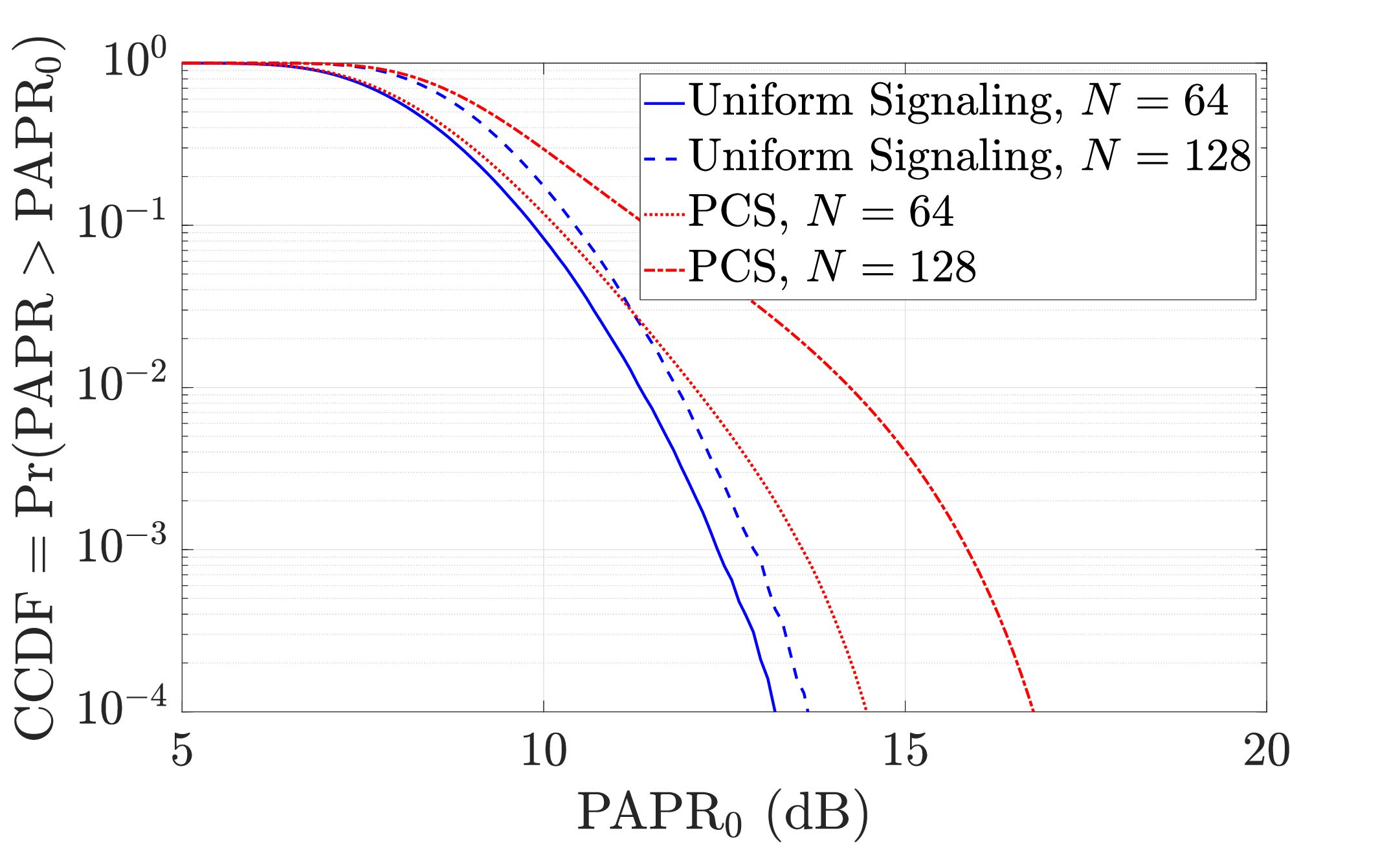}
        \caption{8-QAM.}
        \label{fig:8QAM}
    \end{subfigure}
    \hfill
    \begin{subfigure}{0.24\textwidth}
        \centering
        \includegraphics[width=\linewidth,height = 0.7\linewidth]{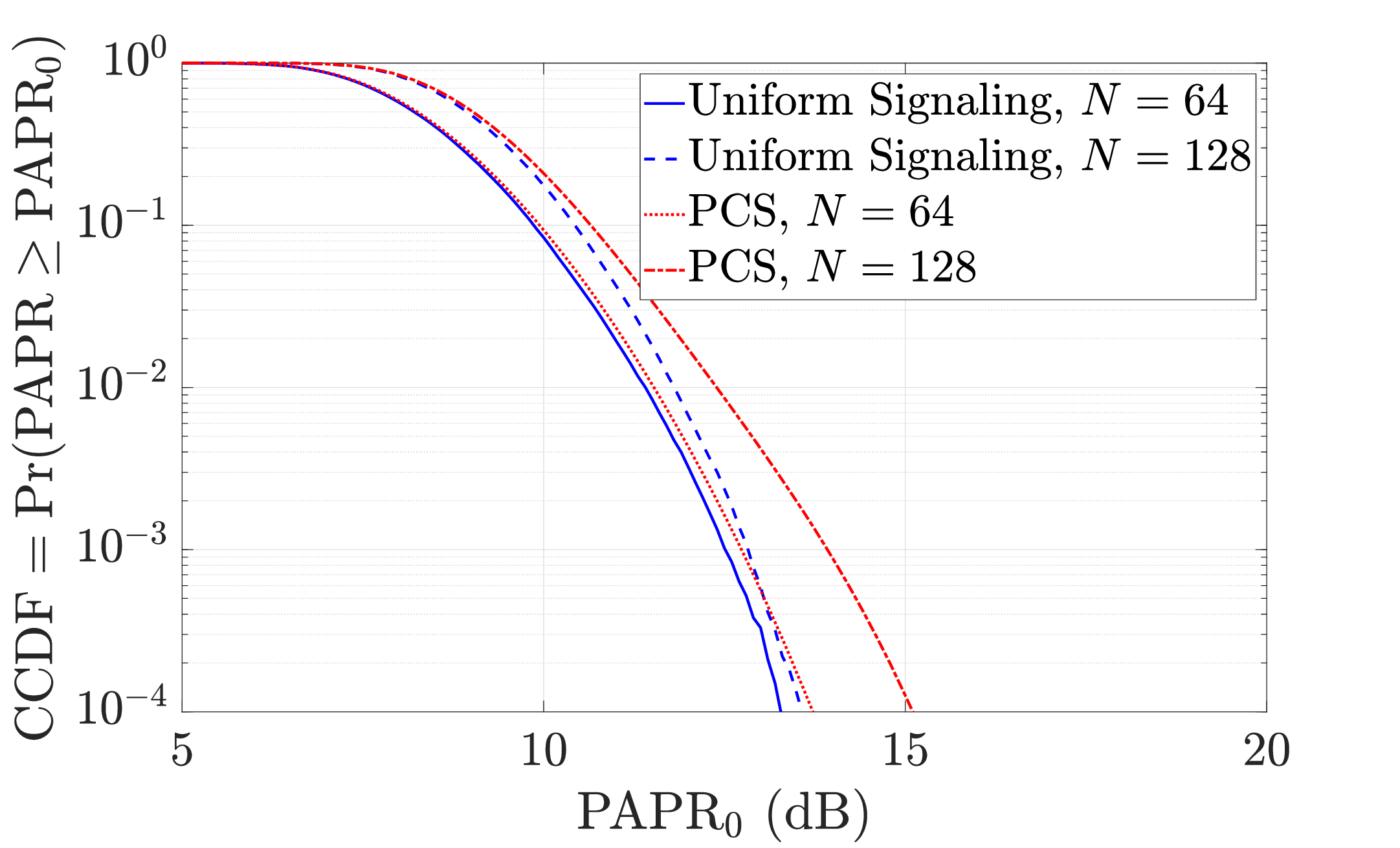}
        \caption{16-QAM.}
        \label{fig:16QAM}
    \end{subfigure}
    \begin{subfigure}{0.24\textwidth}
        \centering
        \includegraphics[width=\linewidth,height = 0.7\linewidth]{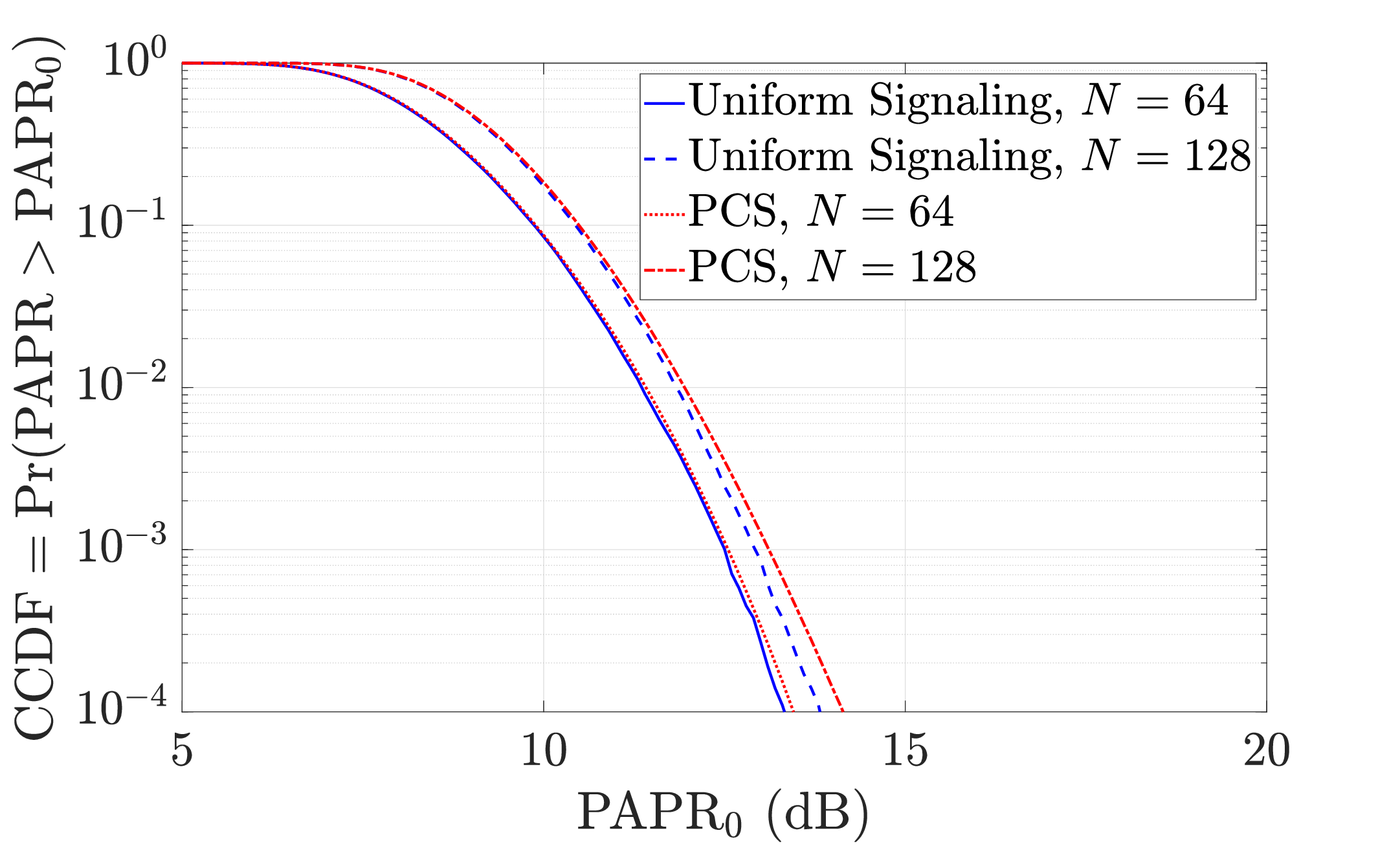}
        \caption{32-QAM.}
        \label{fig:32QAM}
    \end{subfigure}
    \caption{CCDF of PAPR with uniform signaling and PCS.}
    \label{fig:constellation}
    \vspace{-.5cm}
\end{figure}
\vspace{-0.2cm}
\subsection{Clipping Distortion}
Let $I_{\text{min}}$ and $I_{\text{max}}$ be the lower and upper limits of the dynamic linear range of the optical transmitter. Also, denote $I_{\text{DC}}$ as the DC bias. Note that the clipping of $x[n]$ is performed before DC biasing; thus, it is expressed by
\begin{align}
    \tilde{x}[n] = \begin{cases}
    I_{\text{min}} - I_{\text{DC}}, & \text{if $x[n] < I_{\text{min}}$},\\
    x[n] - I_{\text{DC}}, & \text{if $I_{\text{min}} \leq x[n] \leq I_{\text{max}}$}, \\
    I_{\text{max}} - I_{\text{DC}}, & \text{if $x[n] > I_{\text{max}}$}.
  \end{cases}
\end{align}
Let $X_0$, $X_1$, $\dots$, $X_{M-1}$ be the $M$ symbols of the QAM constellation where the occurrence probability of $m$-th symbol $X_m$ is $p_m$ (i.e., $\text{Pr}(X = X_m) = p_m$). Assuming a sufficiently large number of subcarriers (i.e., $N \geq 64$), due to the central limit theorem (CLT), $x[n]$ can be well approximated by a zero-mean Gaussian distribution. The average electrical signal power is thus
\begin{align}
    \sigma^2_x = \left(1 - \frac{2}{N}\right)\sum_{m = 0}^{M-1}p_m\big|X_m\big|^2.
    \label{signal-variance}
\end{align}
According to the Bussgang theorem, the clipped signal $\tilde{x}$ can be written as
\begin{align}
    \tilde{x}[n] = Rx[n] + z_{\text{clip}}[n],
\end{align}
where $R$ is the attenuation factor and $z_{\text{clip}}[n]$ is the additive clipping noise. If we denote $\alpha = \frac{I_{\text{min}} - I_{\text{DC}}}{\sigma_x}$ and $\beta = \frac{I_{\text{max}} - I_{\text{DC}}}{\sigma_x}$, the attenuation factor is given by $R = \mathcal{Q}(\alpha) - \mathcal{Q}(\beta)$ \cite{Dimitrov2012}, where $\mathcal{Q}(x) = \frac{1}{\sqrt{2\pi}}\int_x^{\infty}\exp\left(\frac{-t^2}{2}\right)dt$ is the Q-function. 

Let $\eta$, $\gamma$, and $h$ be the electrical-to-optical conversion factor of the optical transmitter, the responsivity of the PD, and the optical channel gain, respectively. The received time-domain electrical signal is thus given by
\begin{align}
    y[n] = \eta \gamma h(Rx[n] + z_{\text{clip}}[n] + I_{\text{DC}}) + z[n], 
\end{align}
where $I_{\text{DC}}$ is the DC-bias and $z[n]$ denotes the receiver noise, which is modeled as a zero-mean Gaussian distribution with variance $\sigma^2_z = BN_0$. Here, $B$ is the modulation bandwidth and $N_0$ is the noise power spectral density. After the A/D conversion, CP removal, and S/P processing, the time-domain signal $y[n]$ is transformed into its frequency domain through an FFT. Note that the DC-bias term is filtered out since it does not carry information. 
The frequency-domain signal is thus given by
\begin{align}
    Y[k] = \eta\gamma h \big(RX[k] + Z_{\text{clip}}[k]\big) + Z[k],
    \label{sub-channel}
\end{align}
where $Z_{\text{clip}}[k] = \text{FFT}\{z_{\text{clip}}[n]\} \sim \mathcal{CN}(0, \sigma_{\text{clip}}^2)$  and $Z[k] = \text{FFT}\{z[n]\} \sim \mathcal{CN}(0, \sigma^2_z)$. Given a sufficiently large number of subcarriers, $Z_c[k]$ can be approximated as a zero-mean Gaussian random variable according to the CLT. The variance of $Z_c[k]$ is given by \cite{Dimitrov2012}
\begin{align}
    \sigma^2_{\text{clip}} = \sigma^2_x   &\Big(R+\alpha\phi(\alpha)-\beta\phi(\beta)  +\alpha^2(1-\mathcal{Q}(\alpha))+\beta^2\mathcal{Q}(\beta) \big. \nonumber \\ & \left. -\big(\phi(\alpha) - \phi(\beta) + (1-\mathcal{Q}(\alpha))\alpha + \mathcal{Q}(\beta)\beta\big)^2 \right. \!\!\! -  R^2\Big),
    \label{clipping-noise-variance}
\end{align}
where $\phi(t) = \frac{1}{2\pi}\exp\left(\frac{-t^2}{2}\right)$.
\section{Constellation Optimization}
For conciseness, denote $\rho = \eta\gamma h$. The channel capacity of the $k$-th subcarrier given in \eqref{sub-channel} as a function of the symbol distribution $\mathbf{p} = \begin{bmatrix}
    p_0 & p_1 & \cdots \ & p_{M-1}
\end{bmatrix}^T$ is given by
\begin{align}
    C(\mathbf{p}) & = I(X; Y) = h(Y) - h(Y|X) \\
    & = -\int_{\mathbb{C}}p(Y)\log p(Y)dY - h(\rho Z_{\text{clip}} + Z),
    \label{channel-capacity-complex}
\end{align}
where $p(Y)$ denotes the probability density function (PDF) of the complex variable $Y$. Since $Z_{\text{clip}} \sim \mathcal{CN}(0, \sigma^2_{\text{clip}})$, one has $h(\rho Z_{\text{clip}} + Z) = \log\left(\pi e\left(\rho^2\sigma^2_{\text{clip}} + \sigma^2_z\right)\right)$. The PDF $p(Y)$ can be written as
\begin{align}
    p(Y) &= \sum_{m=0}^{M-1}p(Y|X = X_m)\text{Pr}(X = X_m) \nonumber \\
    & = \sum_{m=0}^{M-1}p_m\frac{1}{\pi\left(\rho^2\sigma^2_{\text{clip}} + \sigma^2_z\right)}\exp\left(-\frac{\left| Y - \rho RX_m\right|^2}{\rho^2\sigma^2_{\text{clip}} + \sigma^2_z}\right).
\end{align}
By rewriting  $p(Y)$ as in \eqref{probablity-real-domain}, which is on top of this page, where $Y_r = \mathfrak{R}(Y)$, $Y_i = \mathfrak{I}(Y_i)$ are the real and imaginary parts of $Y_i$ and  $X_{m, r} = \mathfrak{R}(X_m)$, $X_{m, i} = \mathfrak{I}(X_m)$ are the real and imaginary parts of $X_m$, one can transform the complex integration \eqref{channel-capacity-complex} into an integration over the real domain as follows
\begin{figure*}
\begin{align}
p(Y) & = p\left(Y_r + iY_i\right)  = \sum_{m = 1}^M p_m\frac{1}{\pi\left(\rho^2\sigma^2_{\text{clip}} + \sigma^2_z\right)}\exp\left(-\frac{\left(Y_r - RX_{m, r}\right)^2 + \left(Y_i - RX_{m, i}\right)^2}{\rho^2\sigma^2_{\text{clip}} + \sigma^2_z}\right).
\label{probablity-real-domain}
\end{align}
\rule{\linewidth}{0.4pt}
\vspace{-0.9cm}
\end{figure*}
\begin{align}
    C(\mathbf{p})  = & -\!\!\iint_{\mathbb{R}^2}p\left(Y_r \! + \! iY_i\right)\log p\left(Y_r + iY_i\right)dY_rdY_i \nonumber \\
    & - \log\left(\pi e \left(\rho^2\sigma^2_{\text{clip}} + \sigma^2_z\right)\right).
\end{align}
Given an average symbol power limit $P$, we are interested in maximizing the capacity $C(\mathbf{p})$, which is thus formulated as 
\begin{subequations}
\label{OptProb1}
    \begin{alignat}{2}
        &\underset{\mathbf{p}}{\text{maximize}} & \hspace{2mm} & C(\mathbf{p}) \label{obj1}\\
        &\text{subject to }  
        & & \mathbf{a}^T\mathbf{p} \leq P, \label{constraint11}  \\ 
        & & & \mathbf{1}^T_M \mathbf{p} = 1, 
        \label{constraint12} \\
        & & & 0 \leq \mathbf{p} \leq 1, \label{constraint13}
    \end{alignat}
\end{subequations}
where $\mathbf{a} = \begin{bmatrix}\left|X_0\right|^2 & \left|X_1\right|^2 & \cdots & \left|X_{M-1}\right|^2\end{bmatrix}$, $\mathbf{1}_M$ denotes the all-one column vector of size $M$ and $(\cdot)^T$ is the transpose operation. 
Optimally solving \eqref{OptProb1} can be challenging and computationally expensive due to the complex nonlinear nonconvex terms $R$ and $\sigma^2_{\text{clip}}$ that make the objective in \eqref{obj1} a nonconvex function of $\mathbf{p}$. 
In wireless environments where the channel condition often changes quickly, obtaining a PCS (not necessarily optimal) solution within a reasonable amount of time is crucial to enabling real-time transmission. This is especially important in the case of high-order QAM (i.e., larger size of the optimization variable $\mathbf{p}$).
To strive for a balance between quality and complexity, we resort to the simple PGD method for solving \eqref{OptProb1} \cite{boyd2004convex}. The adoption of PGD is also motivated by the fact that the feasible set of \eqref{OptProb1} is convex, which enables efficient handling of the projection step.  The algorithm is outlined as in \textbf{Algorithm 1}.
\vspace{-0.5cm}
\begin{figure}[ht]
\centering
\resizebox{0.48\textwidth}{!}{  
\begin{minipage}{.55\textwidth}
\begin{algorithm}[H]
\caption{PGD method for solving \eqref{OptProb1}}
\begin{algorithmic}[1]
\State Choose a starting point $\mathbf{p}^{(0)}$, a maximum number of iterations $L_{\text{max}}$, a step size $\tau$, and an error tolerance $\varepsilon$.
\For{$k$ = 1 to $L_{\text{max}}$}
    \State Compute the gradient of $C(\mathbf{p})$ at 
    $\mathbf{p}^{(k)}$, denoted as 
    \Statex ~~~~$\nabla C\left(\mathbf{p}^{(k)}\right)$.
    \State Update $\mathbf{q}^{(k+1)} \leftarrow \mathbf{p}^{(k)} - \tau \nabla C\left(\mathbf{p}^{(k)}\right)$. 
    \State Project $\mathbf{q}^{(k+1)}$ onto the feasible set $\mathcal{S}$ of $\eqref{OptProb1}$, i.e., 
    \Statex ~~~~$\mathbf{p}^{(k+1)} = \text{arg}~\underset{\mathbf{p} \in \mathcal{S}}{\text{min}} \left\lVert \mathbf{p} - \mathbf{q}^{(k+1)} \right\rVert$.
    \If{$\frac{\left\lVert\mathbf{p}^{(k+1)} - \mathbf{p}^{(k)}\right\rVert}{\left\lVert\mathbf{p}^{(k)}\right\rVert} \leq \varepsilon$}
        \State \Return $\mathbf{p}^{(k+1)}$
    \EndIf
\EndFor
\end{algorithmic}
\end{algorithm}
\end{minipage}
}
\end{figure}

Note that due to the complexity of the objective function \eqref{obj1}, there is no simple analytical expression for the gradient of $C(\mathbf{p})$. The computation of $\nabla C\left(\mathbf{p}^{(k)}\right)$ is, therefore, performed numerically. The projection of $\mathbf{q}^{(k+1)}$ onto the feasible set of \eqref{OptProb1} can be formulated as the following  problem 
\begin{subequations}
\label{OptProb2}
    \begin{alignat}{2}
        &\underset{\mathbf{p}}{\text{minimize}} & \hspace{2mm} & \frac{1}{2}\left\lVert\mathbf{p} - \mathbf{q}^{(k+1)}\right\rVert \label{obj2}\\
        &\text{subject to } 
        & & \eqref{constraint11},~\eqref{constraint12}, ~\eqref{constraint13} \nonumber.
    \end{alignat}
\end{subequations}
Although several generic solvers (e.g., SDPT3 and SeDuMi) can be used to solve \eqref{OptProb2}, they are designed for tackling general optimization problems, and thus might not be efficient for our specific problem. In the following, we present a nested bisection method to efficiently solve \eqref{OptProb2}.

The Lagrangian of \eqref{OptProb2} is given by
$
  \mathcal{L}(\mathbf{p}, \lambda, \nu) = \frac{1}{2} \left\|\mathbf{p} - \mathbf{q}^{(k+1)}\right\|^2 \! + \! \lambda \left(\mathbf{a}^T \mathbf{p} - P \right) \! + \! \nu\left(\mathbf{1}^T_M\mathbf{p} - 1 \right),  
$
where $\lambda \geq 0$ and $\nu$ are dual variables associated with the constraints \eqref{constraint11} and \eqref{constraint12}, respectively. 
Taking the derivative of the Lagrangian with respect to $\mathbf{p}$  and setting it to zero yields 
$\mathbf{p} = \mathbf{q}^{(k+1)} - \lambda \mathbf{a} - \nu$.
To satisfy the box constraint in \eqref{constraint13}, $\mathbf{p}$ is projected onto the interval $[0,1]$, resulting in $\mathbf{p} = \min\left(1, \max\left(0, \mathbf{q}^{(k+1)} - \lambda \mathbf{a} - \nu \right)\right).$ Then we employ a nested bisection algorithm to find $\lambda$ and $\nu$ given constraints \eqref{constraint11} and \eqref{constraint12}. Specifically, the outer loop adjusts $\nu$ to enforce $\mathbf{1}_M^T\mathbf{p} = 1$ while the inner loop solves for $\lambda$ such that $\mathbf{a}^T \mathbf{p} \leq P$. The algorithm is described in \textbf{Algorithm~\ref{alg:proj}}.
\vspace{-0.2cm}
\begin{figure}[ht]
\centering
\resizebox{0.48\textwidth}{!}{  
\begin{minipage}{.55\textwidth}
\begin{algorithm}[H]
\caption{Nested bisection method for solving \eqref{OptProb2}}
\label{alg:proj}
\begin{algorithmic}[1]
\State Choose an error tolerance $\varepsilon$, a max iterations $L_{\text{max}}$, and a sufficiently large $W$.
\State Define function \texttt{SolveLambda}($\nu$):
\State Initialize $\lambda_{\text{low}} \gets 0$, $\lambda_{\text{high}} \gets W$
\For{$l_1 = 1$ to $L_{\text{max}}$}
    \State $\lambda \gets (\lambda_{\text{low}} + \lambda_{\text{high}}) / 2$
    \State $\mathbf{p} \gets \min\left(1, \max\left(0, \mathbf{q}^{(k+1)} - \lambda \mathbf{a} - \nu \right)\right)$
    \If{$\mathbf{a}^T\mathbf{p} - P \leq \varepsilon$ \textbf{or} $\lambda_{\text{high}} - \lambda_{\text{low}} \leq \varepsilon$}
        \State \Return $\mathbf{p}$
    \ElsIf{$\mathbf{a}^T\mathbf{p} > P$}
        \State $\lambda_{\text{low}} \gets \lambda$
    \Else
        \State $\lambda_{\text{high}} \gets \lambda$
    \EndIf
\EndFor

\State Initialize $\nu_{\text{low}} \gets -W$, $\nu_{\text{high}} \gets W$
\For{$l_2 = 1$ to $L_{\text{max}}$}
    \State $\nu \gets (\nu_{\text{low}} + \nu_{\text{high}})/2$
    \State $\mathbf{p} \gets \texttt{SolveLambda}(\nu)$
    \If{$|\mathbf{1}^T_M\mathbf{p} - 1| \leq \varepsilon$ \textbf{or} $\nu_{\text{high}} - \nu_{\text{low}} \leq \varepsilon$}
        \State \Return $\mathbf{p}$
    \ElsIf{$\mathbf{1}^T_M\mathbf{p} > 1$}
        \State $\nu_{\text{low}} \gets \nu$
    \Else
        \State $\nu_{\text{high}} \gets \nu$
    \EndIf
\EndFor
\end{algorithmic}
\end{algorithm}
\end{minipage}
}
\end{figure}
\section{Simulation Results and Discussions}
We consider a Luxeon\textsuperscript{\textregistered} Rebel LED whose linear dynamic range is $I_{\text{min}} = 100$ mA and $I_{\text{max}} = 1000$ mA \cite{luxeon}. Other system parameters are chosen as follows: DC-bias $I_{\text{DC}} = 500$ mA, modulation bandwidth $B = 20$ MHz, noise power spectral density $N_0 = 10^{-16}$ $\text{mA}^2/\text{Hz}$, number of subcarriers $N = 128$, the electrical-to-optical conversion factor $\eta = 0.44$ W/A, the PD responsivity $\gamma = 0.54$ A/W, and the channel gain $h = 3\times10^{-6}$. For $\textbf{Algorithm 1}$ and $\textbf{Algorithm 2}$, $L_{\text{max}} = 500$, $\varepsilon = 10^{-3}$, $\tau = 10^{-4}$, and $W = 10^{5}$ are set.  

\begin{figure}[t]
    \vspace{-0.4cm}
    \centering
    \includegraphics[width=0.75\linewidth, height = .55\linewidth]{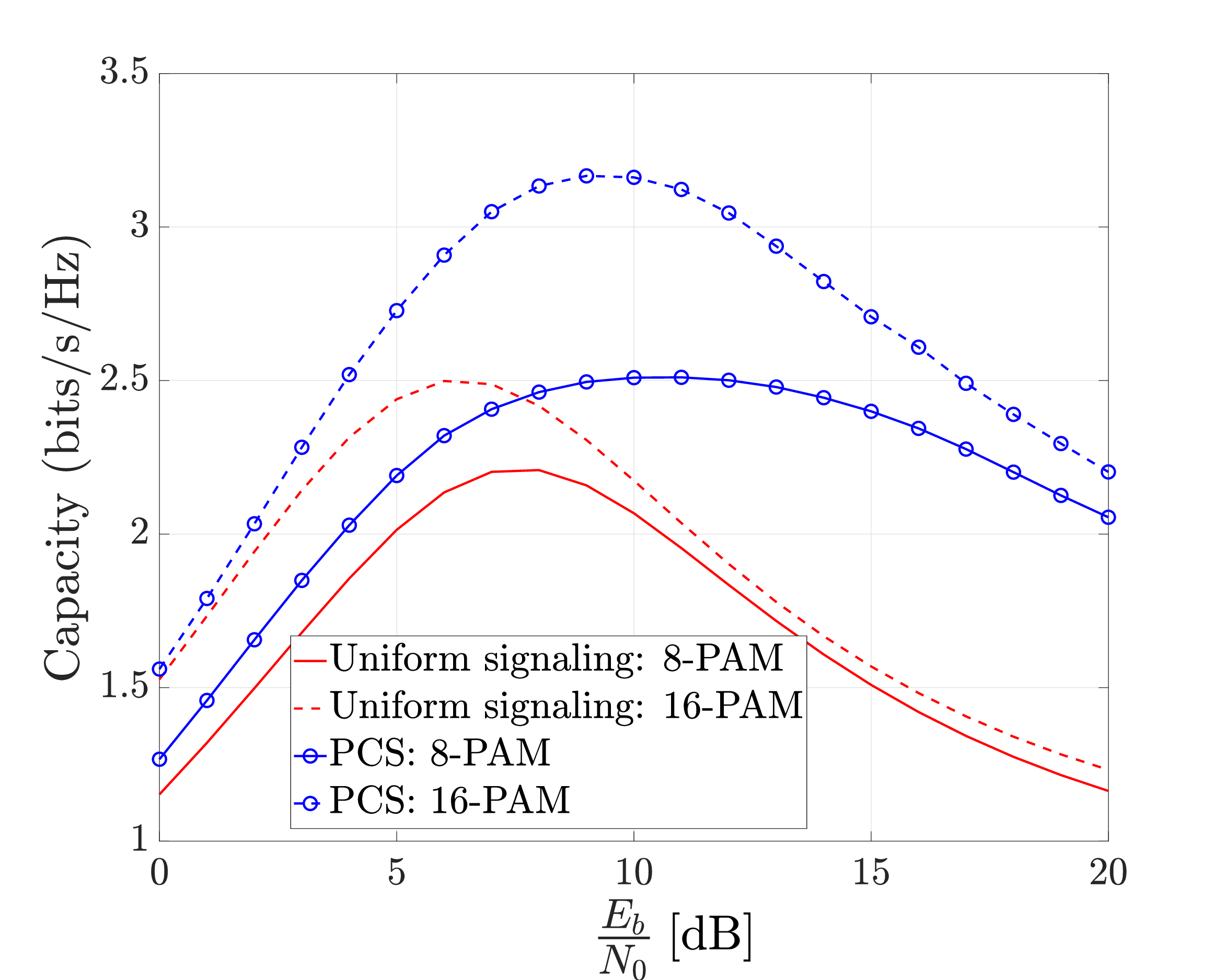}
    \caption{Capacities of PCS and uniform signaling.}
    \label{fig:capacity}
    \vspace{-0.4cm}
\end{figure}

Firstly, we show in Fig.~\ref{fig:capacity} capacity comparisons between PCS and uniform signaling for 8-PAM and 16-PAM constellations. The capacities are plotted in accordance with the average bit energy to noise power spectral density ratio $\frac{E_b}{N_0}$, which, for DCO-OFDM systems, is given by $\frac{E_b}{N_0} = \frac{\sigma^2_x}{\log2(M)\frac{N-2}{N}BN_0}$ \cite{Dimitrov2012}. It is seen that there exist optimal $\frac{E_b}{N_0}$ values beyond which the capacities decreases as $\frac{E_b}{N_0}$ increases. This is due to the fact that increasing $\frac{E_b}{N_0}$ results in a higher variance of the OFDM signal's amplitude, hence a more severe clipping distortion. When $\frac{E_b}{N_0}$ exceeds a certain threshold (i.e., the optimal value), the clipping distortion becomes dominant that increasing $\frac{E_b}{N_0}$ only leads to reduced capacity. The superiority of PCS over uniform signaling is also verified, especially in the clipping distortion-dominant regime. For example, at $\frac{E_b}{N_0} = 15$ dB, capacities of PCS are approximately 59\% and 72.6\% higher than those of uniform signaling in the case of 8-PAM and 16-QAM, respectively. This highlights the benefit of employing PCS for optical OFDM systems under clipping distortion.        

\begin{figure}[ht]
    \vspace{-.5cm}
     \centering
     \begin{subfigure}[b]{0.49\linewidth}
         \centering
         \includegraphics[width=\textwidth]{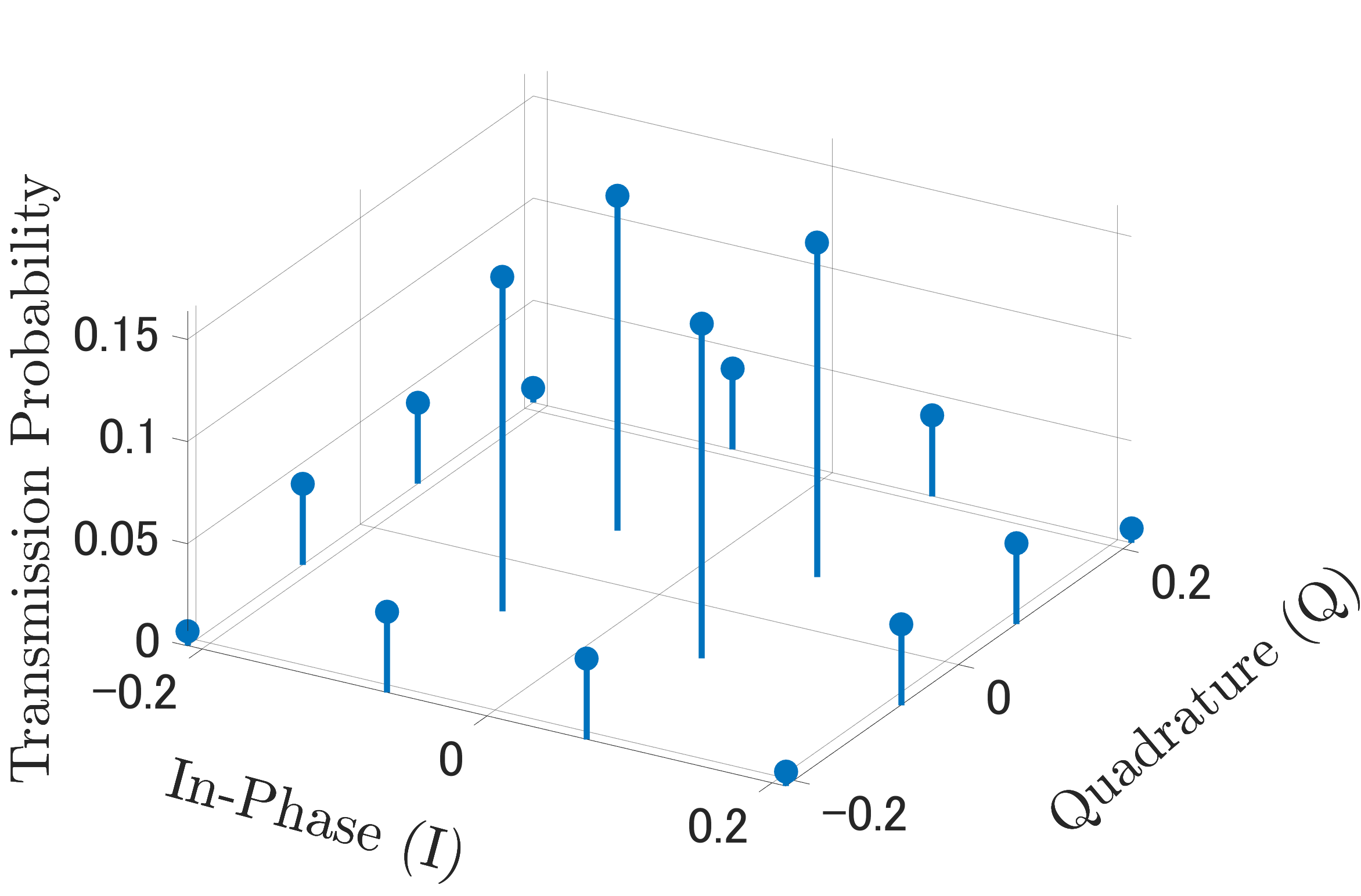}
         \caption{$\frac{E_b}{N_0} = 5$ dB.}
         \label{fig:constellation-1}
     \end{subfigure}
     \begin{subfigure}[b]{0.49\linewidth}
         \centering
         \includegraphics[width=\textwidth]{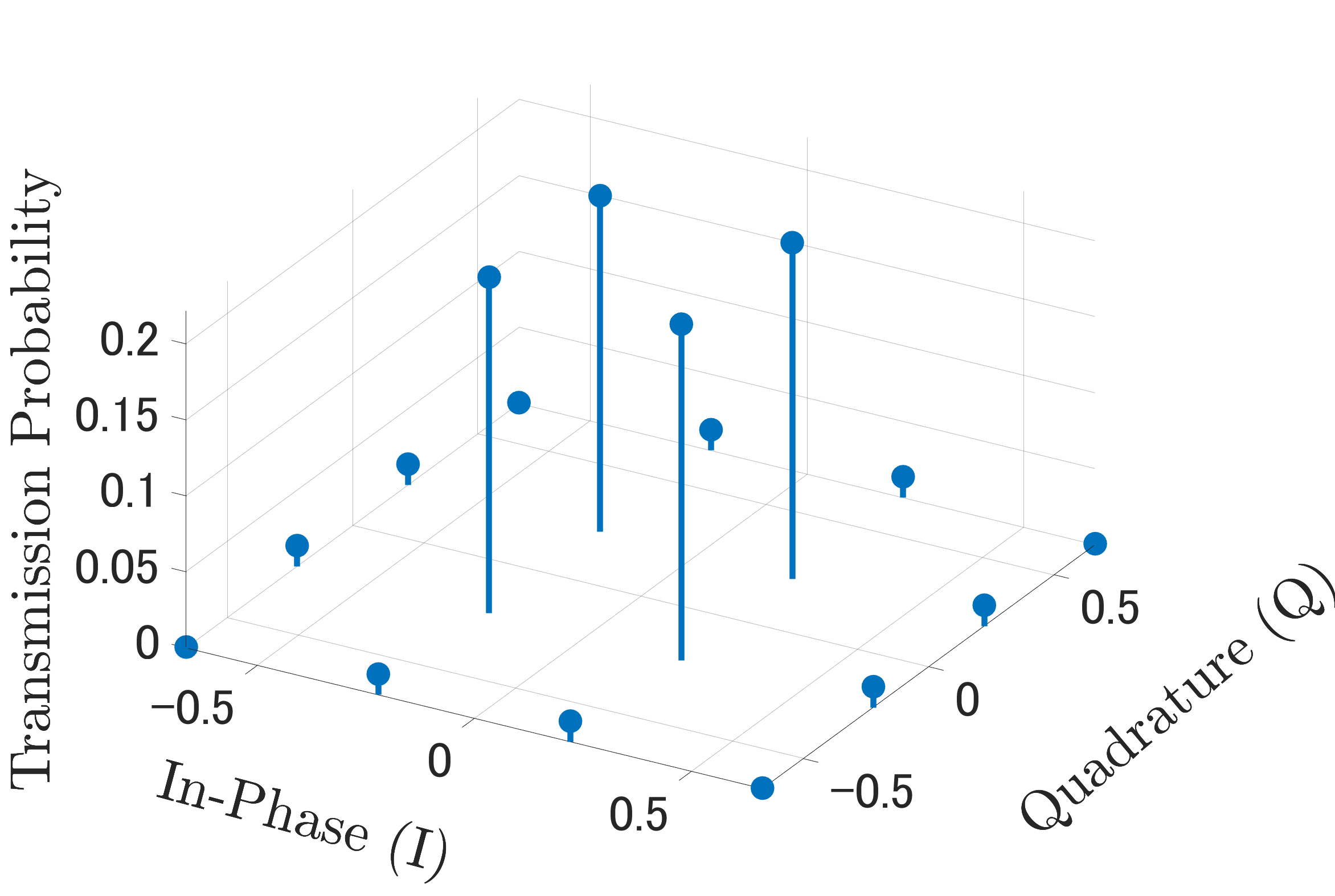}
         \caption{$\frac{E_b}{N_0} = 15$ dB.}
         \label{fig:constellation-2}
     \end{subfigure}
    \caption{Optimal 16-QAM constellations.}
    \label{fig:constellation}
\end{figure}

The optimal 16-QAM constellations are displayed in Figs.~\ref{fig:constellation-1} and \ref{fig:constellation-2} given that  $\frac{E_b}{N_0} = 5$ dB and $15$ dB, respectively. We first notice that the transmission probabilities of symbols with lower energy (i.e., inner points of the constellation) are higher than those of symbols with higher energy. This is expected since low-energy symbols contribute less to clipping distortion. Moreover, when $\frac{E_b}{N_0}$ increases, to combat the clipping distortion, the low-energy symbols are transmitted with even higher probabilities.   

\begin{figure}[t]
    \vspace{-0.4cm}
    \centering
    \includegraphics[width=0.75\linewidth, height = .55\linewidth]{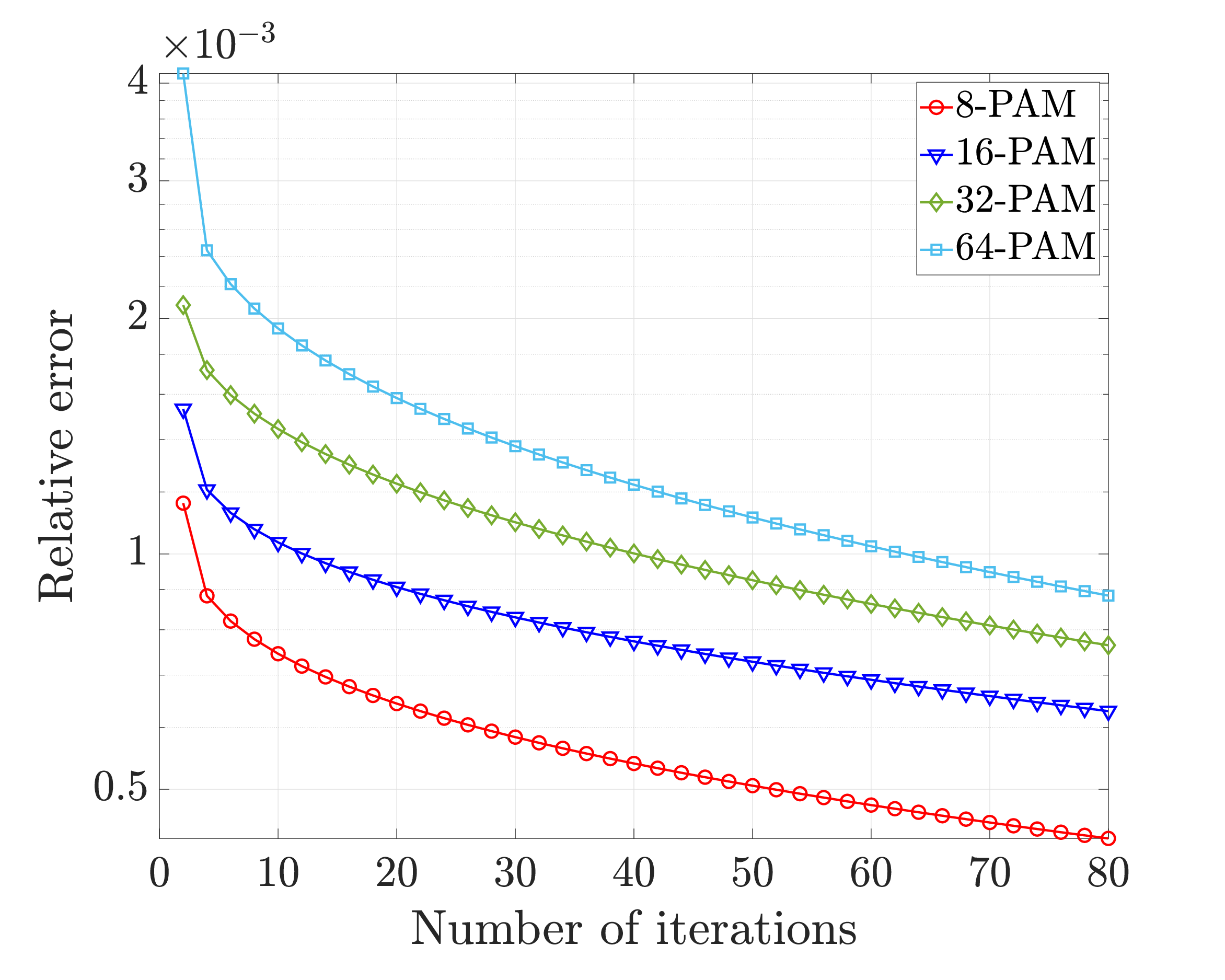}
    \caption{Convergence properties of \textbf{Algorithm 1}.}
    \label{fig:convergence}
\end{figure}
Convergence behaviors of \textbf{Algorithm 1} are illustrated in Fig.~\ref{fig:convergence} for different modulation orders with $\frac{E_b}{N_0} = 5$ dB. The results are obtained by averaging over $100$ different starting points $\mathbf{p}_0$. To meet the error tolerance of $10^{-3}$, the average numbers of iterations are 3, 14, 42, and 64 for 8-, 16-, 32-, and 64-PAM, respectively. To further demonstrate the practicality of the algorithm, we tabulate in TABLE I the average computational times (in seconds) of the projection step in each iteration (the dominant time-consuming task in each iteration of \textbf{Algorithm 1}) using the standard  SPDT3 solver \cite{Toh1999} and our proposed \textbf{Algorithm 2}. Implementations of the two approaches are done using MATLAB R2022b on a Windows 10 desktop computer with an Intel\textsuperscript{\textregistered} Core\textsuperscript{TM} i9-12900 processor. It is clearly shown that the proposed algorithm solves the projection in each iteration much faster than the SDPT3 solver does. For instance, in the case of 64-PAM, each iteration of \textbf{Algorithm 1} is completed in $1.96\times 10^{-4}$ seconds on average, meaning that only $0.0125$ seconds are needed for \textbf{Algorithm 1} to obtain the solution. 
\begin{table}[h]
\centering
\caption{Computational times of SDPT3 and \textbf{Algorithm 2}.}
\begin{tabular}{|c|c|c|}
\hline
Modulation order & SDPT3 & \textbf{Algorithm 2} \\
\hline
$M = 8$ & 0.2318 & $7.18\times 10^{-5}$ \\ \hline
$M = 16$ & 0.2322 & $9.9\times 10^{-5}$ \\ \hline
$M = 32$ & 0.2361 & $1.19\times 10^{-4}$ \\ \hline
$M = 64$ & 0.2451 & $1.96\times 10^{-4}$ \\ 
\hline
\end{tabular}
\end{table}
\vspace{-0.2cm}
\section{Conclusions}
This work investigated an optimization of PCS for DCO-OFDM systems considering the impact of clipping distortion. Simulation results demonstrated that the proposed signal design outperformed the conventional uniform signaling, particularly at the power regime where clipping distortion is severe. The efficiency of the proposed algorithms to solve the optimal constellation in real-time was also experimentally validated. As GCS has been shown to be beneficial for PAPR reduction, one may consider a joint optimization of the PCS and GCS to further alleviate clipping distortion.
\bibliographystyle{ieeetr}
\vspace{-0.4cm}
\bibliography{references}

\end{document}